\documentclass[aps,prc,twocolumn,showpacs,showkeys,preprintnumbers,amsmath,amssymb,a4paper]{revtex4}

\usepackage{epsfig}

\def\lsim{\mathrel{\rlap{\lower4pt\hbox{\hskip1pt$\sim$}}
    \raise1pt\hbox{$<$}}}                
\def\gsim{\mathrel{\rlap{\lower4pt\hbox{\hskip1pt$\sim$}}
    \raise1pt\hbox{$>$}}}                
\newcommand{\bVec}[1]{\mbox{\boldmath$#1$}}

\begin{document}
\title{Precise calculation of the two-step process for $K^- d \to \pi \Sigma N$
in the $\Lambda$(1405) resonance region}

\author{K. Miyagawa$^{1}$ and J. Haidenbauer$^{2,3}$}

\affiliation{
$^{1}$Simulation Science Center, Okayama University of Science, 
1-1 Ridai-cho, Okayama 700-0005, Japan
\\
$^{2}$Institute for Advanced Simulation,
Forschungszentrum J\"ulich, D-52425 J\"ulich, Germany \\
$^{3}$Institut f\"ur Kernphysik and J\"ulich Center for Hadron Physics,
Forschungszentrum J\"ulich, D-52425 J\"ulich, Germany
}

\begin{abstract}
The reaction $K^- d \to \pi \Sigma N$ is investigated taking into account 
single scattering and the two-step process due to $\bar K N \to \pi \Sigma$ 
rescattering. The influence of some common approximations are examined. 
It is found that the treatment of the kinematics 
in the Green's function that appears in the loop integral of
the rescattering process has a rather strong impact on the resulting 
lineshape of the $\pi \Sigma$ invariant mass spectrum. 
Specifically, a calculation with correct kinematics where the 
three-body unitarity cut due to the $nK^- p$ threshold
occurs at the physical value yields a pronounced peak in the 
invariant mass spectrum at this threshold and, at the same time, 
suppresses the signal in the region of the $\Lambda$(1405) resonance. 
On the other hand, an approximation applied in past calculations 
shifts that threshold down and, consequently, leads to an 
accidental and therefore erroneous enhancement of the signal of the 
$\Lambda$(1405) in the $\pi \Sigma$ invariant mass spectrum. 
\end{abstract}

\pacs{13.75.Jz, 12.39.Pn} 
\keywords{ }

\maketitle

\section{Introduction}
\label{sec:intro}

The $\Lambda$(1405), a baryon resonance with 
$I(J^P) = 0(\frac{1}{2}^-)$, has intrigued theorists already
for several decades. The proximity of its nominal mass \cite{PDG}
to the $\bar KN$ threshold (at around 1435 MeV) has led to 
speculations that this resonance is, in fact, a $\bar KN$ (quasi) bound 
state rather than a genuine 3-quark state as soon as it was 
experimentally identified in the early 1960s. A further and even more
peculiar facet was added to this when it was suggested that the 
$\Lambda$(1405) could be actually a superposition of two resonance
states \cite{Oller01,Jido03}. This conjecture emerged from model calculations
performed within the so-called chiral unitary approach based on
coupled channels ($\bar KN$, $\pi\Sigma$, ...). 

Subsequent investigations conducted within variants of that approach,
utilizing the leading-order chiral Lagrangian (Weinberg-Tomozawa term) 
as interaction potential but also higher-order contributions, 
supported the existence of two poles in the energy region of the 
$\Lambda$(1405) resonance \cite{Carmen02,Oller05,Bora,Borasoy:2005ie,Mai12}. 
Thereby it was found that typically one
of the poles lies very close to the $\bar KN$ threshold, i.e. around
1420-1430 MeV, and couples strongly to the $\bar KN$ system \cite{Jido03}.
The other pole exhibits a much larger variation from model to model, 
i.e. is usually located around 1340--1400 MeV (though even values 
around 1470 MeV are reported \cite{Mai12}) and has usually a much larger 
width. Furthermore, it couples more strongly to the $\pi \Sigma$ system. 

Naturally, the prospect of finding two $\Lambda$(1405) resonances has 
trigged also an increased interest in performing corresponding experiments. 
These experiments are guided by the idea that reactions that are dominated
by either the $\bar KN$ or the $\pi \Sigma$ transition channels should
then also provide evidence for the presence of either the one or the other
corresponding pole. Specifically, $K^-$ induced reactions should then be 
dominated by the pole around 1420 MeV and, accordingly, show an enhancement
in the distribution at the corresponding invariant mass \cite{Jido03}. 

In the present work we consider the reaction $K^- d \to \pi \Sigma N$ 
where the $\Lambda(1405)$ can be excited. Our study is motivated by
a corresponding proposal submitted to the J-PARC 50-GeV proton synchrotron.
This proposal aims at a spectroscopic study of hyperon resonances below the $\bar K N$
threshold via the ($K^-,n$) reaction on a deuteron target \cite{jparc}. The
primary goal of the experiment is to study the position and width of the 
$\Lambda$(1405) resonance produced in the $\bar K N \to \pi \Sigma$ channel.
For this reaction theoretical investigations were presented by 
Jido et al. in \cite{Jido09} and, with emphasis on the kinematical 
conditions of the DAFNE facility at Frascati, in \cite{Jido10}. 
Their calculation is performed in impulse approximation and considers
for the reaction mechanism single ($\bar K N \to \pi \Sigma$)
scattering but also the two-step process 
where the kaon first scatters off one of the nucleons and then undergoes 
the transition $\bar K N \to \pi \Sigma$ on the other nucleon.
The required elementary $\bar K N \to \bar K N$ and $\bar K N \to \pi \Sigma$ 
amplitudes are taken from the Oset-Ramos model \cite{OR98,ORB02} that utilizes the 
Weinberg-Tomozawa term as interaction potential.  
The model calculation of \cite{Jido09} yields results that are 
roughly in line with an old measurement of the $\pi^+ \Sigma^-$ 
invariant mass spectrum for the reaction in question from 1977 \cite{Braun}.
Indeed the data exhibit a peak around $M_{\pi\Sigma}\approx 1425$ MeV,
i.e. at roughly the energy where all modern $\bar KN$ interactions cited 
above predict a pole so that everything seemed to match perfectly. 
However, because some approximations are applied in the study of \cite{Jido09} 
this apparent success has to be taken with a grain of salt. 

Our investigation intends to scrutinize the results of Ref.~\cite{Jido09}
in two aspects. First and most importantly, we want to 
avoid some of the approximations introduced in Ref. \cite{Jido09}. 
For example, we do not use factorization, i.e. we do not pull out the 
($\bar K N \to \bar K N$ and $\bar K N \to \pi \Sigma$) amplitudes from the
loop integral that occurs in the calculation of the two-step process.
Also, and more importantly, we treat the kinematics in the Green's function 
that appears in the loop integral properly. Specifically we make sure that
the three-body unitarity cut for the intermediate $\bar K NN$ system
occurs at the correct (physical) threshold. As we will see this has a
decisive influence on the achieved results. 

In addition, we also consider different models for the elementary 
$\bar KN$-$\pi \Sigma$ interaction. Practically all the interactions in the
literature are fitted to the near-threshold cross sections for $\bar K N$ elastic 
and charge-exchange scattering and for the $\bar K N \to \pi\Lambda$ and
$\bar K N \to \pi\Sigma$ transitions. 
As a consequence, the properties of these interactions in the 
$\bar K N$ threshold region are very similar, even down to the position of the
(nominal) $\Lambda$(1405) resonance. However, for energies further away from
the threshold there is a significant model dependence. This is reflected, for
example, in the large variation of the position of the lower pole, already
mentioned above, see also \cite{Ikeda11}. 
Indeed there are phenomenological models that describe 
the data around the $\bar K N$ threshold with comparable quality, but do not
even have a second pole \cite{Revai09}. It is interesting to see whether and how 
these model differences are reflected in the results for $K^- d \to \pi \Sigma N$.
After all, the $\pi \Sigma$ invariant mass spectrum samples the properties of
the $\bar K N$-$\pi \Sigma$ interaction down to the $\pi \Sigma$ threshold. 

In the present study we utilize the Oset-Ramos interaction \cite{OR98,ORB02} so 
that we can compare our results directly with other ones that can be already found 
in the literature \cite{Jido09}. The pole positions produced by this interaction
in the isospin I=0 channel, which are associated with the $\Lambda$(1405),
are 1426 + $i$16 MeV and 1390 + $i$66 MeV \cite{Jido03}, respectively. 
In addition we use a potential model that differs not only in the position of the 
lower pole from the Oset-Ramos interaction \cite{OR98} but also conceptually. 
In particular, we resort to a meson-exchange potential of the 
$\bar K N$-$\pi \Sigma$ systems that was
published by the J\"ulich group more than 20 years ago \cite{MG}, i.e. 
long before the chiral unitary approach became popular. As can be seen in the
original paper \cite{MG}, the J\"ulich model describes the $\bar K N$ scattering 
data in the near-threshold region quite satisfactorily. Other threshold 
quantities are fairly well reproduced too, as shown in a recent paper \cite{Hai11}. 
Of importance for the present study is also that the J\"ulich model
generates likewise two poles in the region of the $\Lambda$(1405) resonance.
One pole, the $\bar KN$ ``bound state'', is located fairly close
to the $\bar KN$ threshold and to the physical real axis 
(1436 + $i$26 MeV) while the other one is close to the $\pi\Sigma$ threshold
and has a significantly larger imaginary part (1334 + $i$62 MeV). In fact,
this pole lies at the lower end of the ''lower pole spectrum'' mentioned above. 

The paper is structured as follows:
In the subsequent section we summarize shortly the salient features of the
$\bar KN$ interaction of the J\"ulich group. 
In Sect.~\ref{sec:formalism} we describe in detail the formalism that is 
employed in our calculation of the reaction $K^- d \to \pi \Sigma N$. 
The results of our calculation for the Oset-Ramos and the J\"ulich 
$\bar KN$ interactions are presented in Sect.~\ref{sec:results}.
In particular, we discuss approximations applied in previous investigations 
and study their impact on the shape of the $\Sigma N$ invariant mass 
spectrum.
The manuscript closes with a summary. 

\section{The J\"ulich $\bar KN$ model}
\label{sec:meson}

The J\"ulich meson-exchange model of the $KN$ and $\bar KN$ 
interactions has been described in detail in the literature 
\cite{Juel1,Juel2,Juel3,Juel4} and we refer the reader to 
those works.  The interaction model was constructed along
the lines of the (full) Bonn $NN$ model \cite{MHE} and its extension
to the hyperon-nucleon ($YN$) system \cite{Holz} ($Y=\Lambda, \ \Sigma$).
Specifically, this means that one
has used the same scheme (time-ordered perturbation theory), the same
type of processes, and vertex parameters (coupling constants, cut-off
masses of the vertex form-factors) fixed already by the study of
these other reactions.

\begin{figure}[t]
\vspace*{+1mm}
\centerline{\hspace*{3mm}
\psfig{file=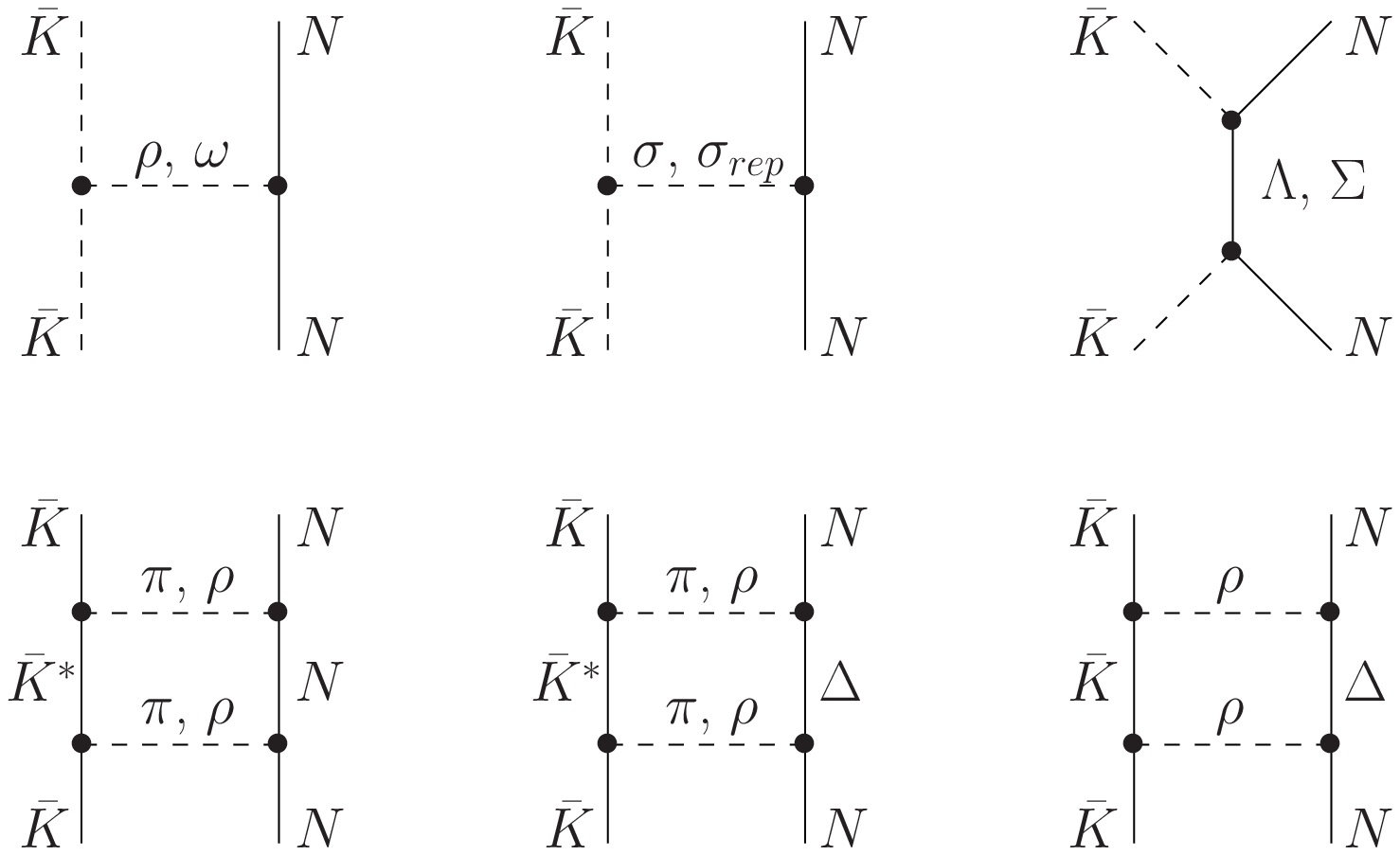,width=11.0cm,height=13.0cm}}
\vspace*{-7.5cm}
\caption{Meson-exchange contributions included in the $\bar K N$
interaction.
}
\label{Diakn}
\end{figure}

The diagrams considered for the $\bar KN$ interaction are shown in
Fig.~\ref{Diakn}. Obviously the J\"ulich model contains not only 
single-meson exchanges, but also higher-order box diagrams
involving $N\bar K^*$, $\Delta \bar K$ and $\Delta \bar K^*$ intermediate
states. Most vertex parameters involving the nucleon and the $\Delta$(1232)
isobar are taken over from the (full) Bonn $NN$ potential.
The coupling constants at vertices involving strange baryons are fixed
from the $YN$ model (model B of Ref.~\cite{Holz}). Those quantities
($g_{N\Lambda K}$, $g_{N\Sigma K}$, $g_{NY^* K}$) have been related to the
empirical $NN\pi$ coupling by the assumption of SU(6) symmetry,
cf. Ref.~\cite{Juel1,Juel2}.

For the vertices involving mesons only, most coupling constants
have been fixed by SU(3) relating them to the empirical $\rho \to
2\pi$ decay. An exception is the coupling constant $g_{KK\sigma}$,
which has been adjusted to the $KN$ data \cite{Juel1}, for
the following reason: The $\sigma$ meson (with a mass of about
600 MeV) is not considered as a genuine particle but as a simple
parametrization of correlated $2\pi$-exchange processes in the
scalar-isoscalar channel. Therefore, its coupling strength cannot
be taken from symmetry relations.
Concerning the $\omega$-exchange the coupling strengths for both 
$g_{NN\omega}$ and $g_{KK\omega}$ were kept at their SU(6) values. 
At the same time a phenomenological, very short-ranged contribution 
was added, denoted as $\sigma_{rep}$. This phenomenological piece has 
the same analytical form as $\sigma$-exchange, but an exchange mass of 
1200 MeV and, most importantly, an opposite sign. 
Such a short-range contribution was required in order to obtain 
sufficient repulsion for a reasonable description of the $S$-wave $KN$ 
phase shifts \cite{Juel1}. It was shown in Ref.~\cite{HHK} that this
phenomenological piece can be explained dynamically, even on a 
quantitative level, by genuine quark-gluon exchange processes. 

The contributions to the $\bar KN$ interaction in \cite{Juel2} are
fixed from those of the $KN$ model \cite{Juel1} via a G-parity 
transformation. 
The only exception is the phenomenological $\sigma_{rep}$ whose
strength is re-adjusted by a fit to $\bar KN$ data. Its contribution 
required there was found to be considerably reduced as compared to
$KN$. Indeed, this is in line with the results of \cite{HHK}
because the quark-gluon exchange processes that generate most of
the repulsion simulated by the $\sigma_{rep}$ in case of $KN$
are absent in the $\bar KN$ channel due to the different quark
structure of the $\bar K$ meson. 
 
\begin{figure}[t]
\vspace*{+1mm}
\centerline{\psfig{file=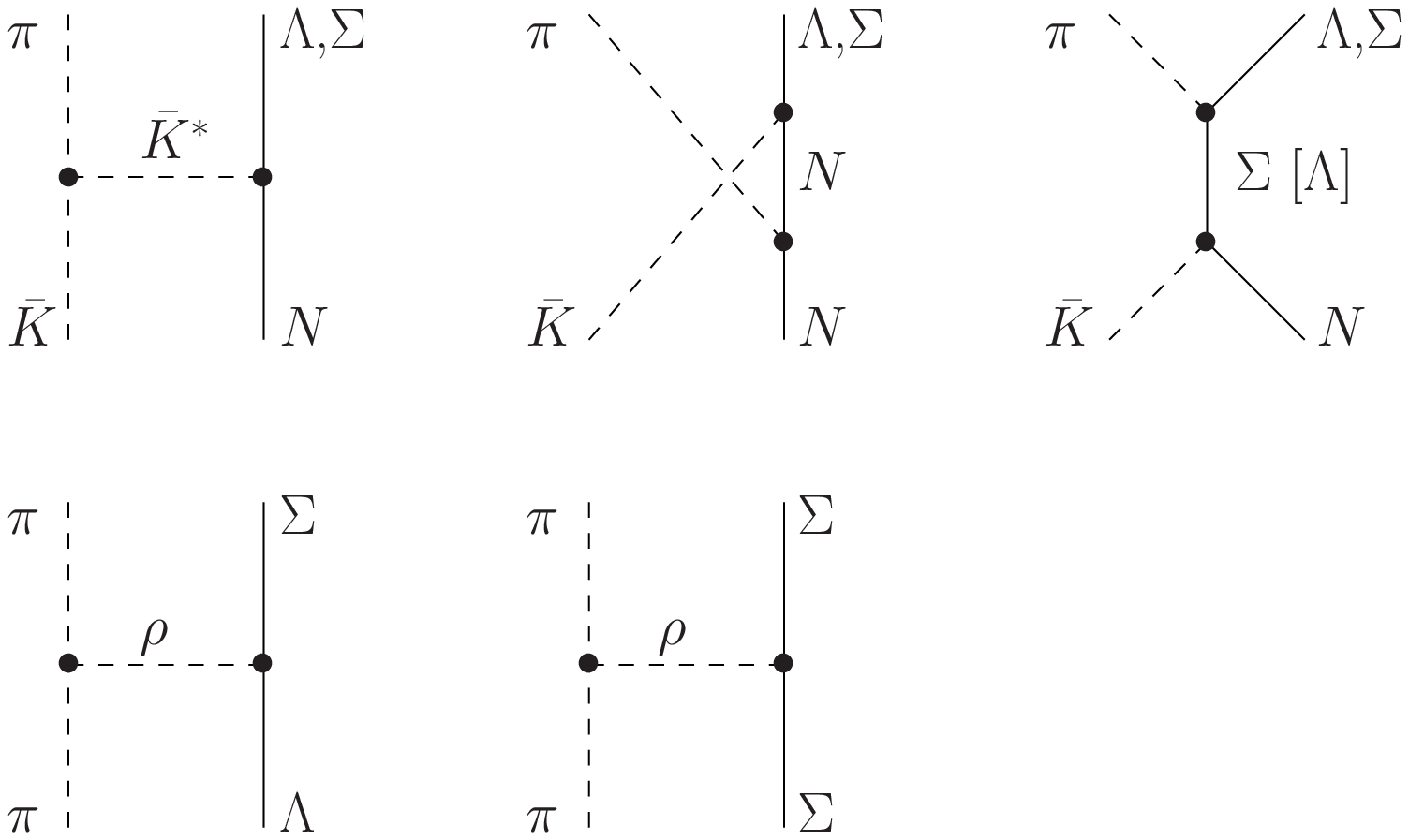,width=11.0cm,height=13.0cm}}
\vspace*{-7.5cm}
\caption{Meson-exchange contributions included in the 
$\bar KN \rightarrow \pi\Lambda, \pi\Sigma$ transition potentials
and in the 
$\pi\Lambda, \pi\Sigma \rightarrow \pi\Lambda, \pi\Sigma$
interactions.
}
\label{Dialp}
\end{figure}

Of course,
in case of the $\bar KN$ system there are already open channels
at the reaction threshold and the coupling to those channels
($\pi\Lambda$, $\pi\Sigma$) is taken into account explicitly. 
The diagrams considered for the $\bar KN\to \pi Y$ transitions
and the $\pi Y \to \pi Y$ interactions
are shown in Fig.~\ref{Dialp}. Also here SU(3) symmetry has been
used for fixing the vertex parameters as far as possible. 

With the $\bar KN$ potential and the 
$\bar KN\to \pi Y$ and the $\pi Y \to \pi Y$ transition
interaction derived from the diagrams in Figs.~\ref{Diakn} 
and \ref{Dialp}, the reaction amplitude $T$ is obtained by a solving a 
(coupled-channels) Lippmann-Schwin\-ger type
equation defined by time-ordered perturbation theory:
\begin{equation}
T_{\alpha\beta} = V_{\alpha\beta} 
+ \sum_\gamma V_{\alpha\gamma} G_{0,\gamma} T_{\gamma\beta} \ 
\label{LSE}
\end{equation}
with $\alpha,\beta,\gamma$ = $\bar KN$, $\pi\Lambda$, $\pi\Sigma$. 

\section{Formulation of {\boldmath $K^- d \to \pi \Sigma n$ } }
\label{sec:formalism}

\begin{figure}[t]
\includegraphics[width=9cm]{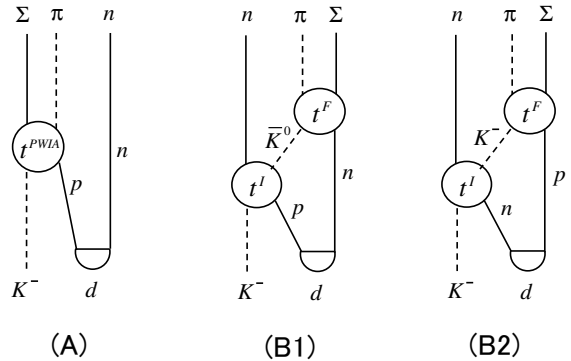}
\caption{Mechanisms included in our calculation of the reaction 
$K^- d \to \pi\Sigma n$. Plane-wave impulse-approximation
(A); $\bar K^0 n \to \pi \Sigma$  (B1) and 
$K^- p \to \pi \Sigma$ (B2) rescattering, respectively. 
}
\label{f4_1}
\end{figure}

\begin{figure}[t]
\includegraphics[width=4cm]{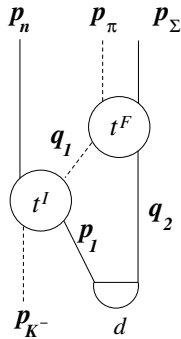}
\caption{Definition of the kinematical variables used in our calculation 
of the reaction $K^- d \to \pi\Sigma n$.
}
\label{f4_2}
\end{figure}

In our study of the reaction $K^- d \to \pi \Sigma n$ we
include the three diagrams shown in Fig.~\ref{f4_1}. Other 2-step 
processes in conjunction with process A in the form of a
subsequent $\pi n$ or $\Sigma n$ final-state interaction (FSI)
are neglected. This is done because, as will be demonstrated later,
the contribution from the process A to the cross section 
is a factor $10^{2}\sim 10^{3}$ smaller than the one from 
process B2 in the considered region of incident $K^-$ lab 
momenta around $p_{k^-}=600$ MeV$/c$.
The general expression of the cross section is given by
\begin{eqnarray}
d\sigma &=&
\frac{1}{|\bVec{v}_{K^-}-\bVec{v}_d |}
(2\pi)^{4}\delta^{4}(p_{n}+p_{\pi}+p_{\Sigma}-p_{K^-}-p_{d})\,
\nonumber\\ &&\times\,
\big\vert
\langle\bVec{p}_n\vert
\langle\bVec{p}_{\pi}\vert
\langle\bVec{p}_{\Sigma}\vert
\, T \,
\vert \bVec{p}_{K^-}\rangle
\vert\Phi_{d}\rangle\,
\big\vert^2
\nonumber\\ &&\times\,
\frac{d^3 p_{n}}{(2\pi)^3} 
\frac{d^3 p_{\pi}}{(2\pi)^3} 
\frac{d^3 p_{\Sigma}}{(2\pi)^3}
\label{F1}
\end{eqnarray}
where the obvious dependence of the cross section on
spin variables is omitted.
The matrix element is given by 
\begin{eqnarray}
\langle\bVec{p}_n\vert &&
\langle\bVec{p}_{\pi}\vert
\langle\bVec{p}_{\Sigma}\vert
 \, T \,
\vert \bVec{p}_{K^-}\rangle
\vert\Phi_{d}\rangle\, 
\nonumber \\
&&=\sqrt{2}\,\,t^{PWIA}(\bVec{p}_{\pi}\,\bVec{p}_{\Sigma}\, ,\,
\bVec{p}_{K^-}\tilde{\bVec{p}}_1)\, \Phi_d(\tilde{\bVec{p}})  
\nonumber \\
&&+\sqrt{2}\int \frac{d^3 q_2}{(2\pi)^3}\,
t^F(\bVec{p}_{\pi}\,\bVec{p}_{\Sigma}\, ,
\bVec{q}_1\bVec{q}_2) \,
G_0 (\bVec{q}_1 \bVec{q}_2)  
\nonumber \\
&&  \hspace{18mm} \times\,
t^I(\bVec{q}_1\,\bVec{p}_n\, ,
\bVec{p}_{K^-}\bVec{p}_1)\, \Phi_d(\bVec{p}) ~. 
\label{F2}
\end{eqnarray}
The first term on the right-hand side is the plane-wave impulse-approximation
(PWIA) which is given by the contribution of diagram A,
while the second term refers to diagram B, whose contribution will be 
discussed and shown explicitly later on for the two possible intermediate particle 
states (B1 and B2). 
The factor $\sqrt{2}$ comes from the proper antisymmetrization.
The quantities $t^F$ and $t^I$ denote the $\bar K N \to \pi \Sigma$
and $\bar K N \to \bar K N$ amplitudes, respectively. 
The various momentum variables which appear in the second term are
depicted in Fig.~\ref{f4_2}.
We work in the lab frame (deuteron rest frame), and then the momenta
in Eq.~(\ref{F2}) satisfy
\begin{eqnarray}
\tilde{\bVec{p}}&=&\tilde{\bVec{p}}_1=-\bVec{p}_n  \: , \nonumber\\
       \bVec{p}&=&        \bVec{p}_1=-\bVec{q}_2 \: , \nonumber\\
      \bVec{q}_1&=& \bVec{p}_1+\bVec{p}_{K^-} -\bVec{p}_n \: . 
\label{F3}
\end{eqnarray}
The meson-baryon two-body energy $E^I$ of the fully off-shell
$t$-matrix $t^I$ is given
by 
\begin{eqnarray}
E^I&=&E_{total}-\sqrt{\bVec{q}_2^2+ m_N^2} \nonumber\\
   &=&E_{K^-}+m_d-\sqrt{\bVec{q}_2^2+ m_N^2} \:\; .
\end{eqnarray}

Equations~(\ref{F1}) and (\ref{F2}), presented here in a rather compact form,  
can be derived within a field theoretical approach in an appropriate manner
(see, for example \cite{BjorkenDrell}).
One only has to take care that the bound state deuteron in the initial state
is incorporated, which should be described as a state in the Heisenberg 
representation (see also, for example \cite{Weinberg}).
Since we will use different interaction models for generating the $t$-matrices,
derived in different frameworks, and a nonrelativistic deuteron wave function,
those equations are not written in invariant form.
Also notice that we assume that the intermediate $\bar{K}$ with momenta $\bVec{q}_1$
propagates forward in time, 
so that the Green's function $G_0$ will be described only by the positive-frequency part. 
This is a quite reasonable treatment because we consider transitions to final states
in the low-energy region around the $n \bar K N$ threshold.

Let us now derive the expression for the inclusive
\noindent
$d(K^-, n)\pi\Sigma$ cross section, where $\pi\Sigma$ indicates one of
the charge states $\pi^+\Sigma^-$, $\pi^0\Sigma^0$, or $\pi^-\Sigma^+$ .
For evaluating the cross section it is convenient to take as integration variables
the direction of the pion momentum $\bVec{p}_\pi^{cm}$ in the 
center-of-mass (c.m.) frame of $\pi$ and $\Sigma$.
Thus, we first rewrite part of the phase space factor 
\begin{eqnarray}
&\delta^{4}&(p_{n}+p_{\pi}+p_{\Sigma}-p_{K^-}-p_{d})\,
d^3 p_{n} \,
d^3 p_{\pi} \,
d^3 p_{\Sigma}\,
\nonumber \\
&=&\delta (E^{cm}_{\pi}+E^{cm}_{\Sigma}-W_{\pi\Sigma})\,
\delta^{3}(\bVec{p}^{cm}_{\pi}+\bVec{p}^{cm}_{\Sigma})\,
\nonumber \\
&&\times
\frac{E_{\pi}}{E^{cm}_{\pi}}\,
\frac{E_{\Sigma}}{E^{cm}_{\Sigma}}\,
d^3 p_{n} \,
d^3 p_{\pi}^{cm}\, 
d^3 p_{\Sigma}^{cm}\,
\label{F4}
\end{eqnarray}
%
where
\begin{eqnarray}
W_{\pi\Sigma}=(E_{K^-}+m_d-E_n)^2-|\bVec{p}_{K^-}-\bVec{p}_n|^2  \: .
\label{F5}
\end{eqnarray}  
Due to the 3-momentum $\delta$-function, the integral over
$\bVec{p}_\Sigma^{cm}$ can be eliminated.
Next, the quantity $d p_\pi^{cm}$ is converted to $d M_{\pi\Sigma}$ 
by the following relation, 
\begin{eqnarray}
d p^{cm}_\pi =\frac{E^{cm}_{\pi}\, E^{cm}_{\Sigma}}
                {M_{\pi\Sigma} \,\, p^{cm}_\pi} d M_{\pi\Sigma} \; ,
\label{F6}
\end{eqnarray}
where
$M_{\pi\Sigma}(= E_\pi^{cm}+E_\Sigma^{cm}$) is
the invariant mass of the $\pi\Sigma$ system.
%
We would like to integrate over the magnitude of the neutron momentum $p_n$, 
which is related to $W_{\pi\Sigma}$ by Eq.~(\ref{F5}). Hence, 
the energy-conserving $\delta$-function is substituted as
\begin{eqnarray}
\delta && (M_{\pi\Sigma} -W_{\pi\Sigma}) \nonumber \\
&&=\frac{W_{\pi\Sigma}}{|(E_{K^-}+m_d)\, {p_n}/{E_n} - p_{K^-} \cos\theta_n|}
\delta (\breve p_n -  p_n) , \qquad
\label{F7}
\end{eqnarray}
where $\theta_n$ is the polar angle of the neutron
with regard to the $K^-$ beam direction, and $\breve p_n$ satisfies 
\begin{eqnarray}
M_{\pi\Sigma}=(E_{K^-}+m_d-\breve E_n)^2-|\bVec{p}_{K^-}-\breve{\bVec{p}}_n|^2
\; .
\label{F8}
\end{eqnarray}  
\noindent
Performing the integral over $p_n$, we obtain the final expression of
the inclusive cross section
\begin{eqnarray}
&&\frac{d \sigma}{d M_{\pi\Sigma} \, d\Omega_n }
\nonumber\\
=&&
\frac{1}{v_{K^-} (2\pi)^5}
\frac{p^{cm}_\pi \, \breve p_n^2}
{|(E_{K^-}+m_d)\, {\breve p_n}/{\breve E_n} - p_{K^-} \cos\theta_n|}
\nonumber\\
&&\times\, \int d\Omega_\pi^{cm}\, E_\pi E_{\Sigma} \,
\big\vert
\langle\breve{\bVec{p}}_n\vert
\langle\bVec{p}_{\pi}\vert
\langle\bVec{p}_{\Sigma}\vert
\, T \,
\vert \bVec{p}_{K^-}\rangle
\vert\Phi_{d}\rangle\,
\big\vert^2 \; . \qquad  
\label{F9}
\end{eqnarray}

Now, let us discuss the second term of the right-hand side of
Eq.~(\ref{F2}) by introducing particle states explicitly. 
This term written out in detail amounts to 
\begin{eqnarray}
\sqrt{2}\:\: \langle n(1) |\, \langle \pi \,\Sigma(2)\, |
\: t^F (2) \, G_0 \, t^I(1)\:
|\Phi_d \rangle |K^- \rangle   
\label{F10}
\end{eqnarray}
where the two baryons are numbered 1 and 2, and the argument 1 in the
operator $t^I (1)$ indicates that it acts only on particle 1.
The same holds for $t^F (2)$.
The process corresponding to $t^F(1)\, G_0\, t^I (2)$ is absorbed into
the factor $\sqrt{2}$.  Applying the operator 
$t^I (1)$ on $|\Phi_d\rangle$ with the isospin part of the deuteron written out
explicitly yields  
\begin{eqnarray}
&& \langle n(1) | \: t^I(1)\: |\Phi_d \rangle |K^- \rangle
\nonumber\\
&& = \langle n(1) | \: t^I(1)\: 
{1\over\sqrt{2}}
[ |\,p(1)\rangle |\,n(2)\rangle -|\,n(1)\rangle |\,p(2)\rangle
 ] \: |\,\phi_d \,\rangle  |K^- \rangle 
\nonumber\\
&& ={1\over\sqrt{2}} [\: 
|\,n(2)\rangle \langle n(1) | \: t^I(1)\: | p(1) K^- \rangle
\nonumber\\
&& \hspace{10mm}
 -|\,p(2)\rangle \langle n(1) | \: t^I(1)\: | n(1) K^- \rangle
\: ] \: |\, \phi_d \,\rangle \; .
\label{F11}
\end{eqnarray}
Inserting the complete set, 
\begin{eqnarray}
|\bar{K^0}n(2) \rangle\langle\bar{K^0}n(2)|+|K^-p(2) \rangle\langle K^-p(2)| , 
\nonumber
\end{eqnarray}
between  $t^F (2) G_0$ and $t^I (1)$ in Eq.~(\ref{F10})
which is allowed by the total-charge conservation,
we end up with 
%
%
%
%
\begin{eqnarray}
&
\sqrt{2} & \:\: \langle n(1) |\, \langle \pi \,\Sigma(2)\, |
\: t^F (2) \, G_0 \, t^I(1)\:
|\Phi_d \rangle |K^- \rangle  
\nonumber\\   
&&=\hspace{2mm}
\langle \pi\Sigma | \: t^F\, G_0\: | \bar{K^0}n\rangle
\, \langle \bar{K^0}n | \: t^I\: | p K^- \rangle
|\,\phi_d \,\rangle 
\nonumber\\
&&\hspace{2mm} -
\langle \pi\Sigma | \: t^F\, G_0\: | K^-p\rangle
\, \langle K^-p | \: t^I\: | n K^- \rangle
|\,\phi_d \,\rangle ,
\label{F12}
\end{eqnarray}
where the first term of the right-hand side corresponds to diagram B1 and the second 
term to diagram B2 in Fig~\ref{f4_1}. Obviously, there is an interference between these 
two terms.

Let us now come to the explicit expression of $G_0(\bVec{q}_1 \bVec{q}_2)$ 
in Eq.~(\ref{F2}). As already mentioned, the 
$\bar{K}$ with momenta $\bVec{q}_1$ propagates forward in time
and $G_0$ is described only by the positive-frequency part. 
Since we work in the lab frame, the Green's function is given by 
\begin{eqnarray}
G_0(\bVec{q}_1 \bVec{q}_2)
&=&\frac{1}{E_1-E_1(q_1)+i\epsilon} 
\nonumber\\
&=&\frac{1}{E_\pi+E_{\Sigma}-E_2(q_2)-E_1(q_1)+i\epsilon} , \quad
\label{F13}
\end{eqnarray}
where
\begin{eqnarray}
E_1&\equiv&E_{total}-E_n-E_2(q_2)  
\nonumber\\
   &=&E_\pi+E_{\Sigma}-E_2(q_2) , 
\label{F131}
\end{eqnarray}
and
\begin{eqnarray}
&&E_2(q_2)=\sqrt{\bVec{q}_2^2+ m_N^2} \;\; ,\quad
E_1(q_1)=\sqrt{\bVec{q}_1^2+ m_{\bar K}^2}  \;\; .\quad
\label{F132}
\end{eqnarray}
The total energy and the energies of the outgoing particles
are indicated by $E_{total}$ and by $E_\pi$, $E_{\Sigma}$, $E_n$, respectively.
We can express the lab energies $E_\pi+E_\Sigma$ and $E_2(q_2)+E_1(q_1)$ 
in Eq.~(\ref{F13}) by using the energies
in the c.m. frame of the $\bar{K} N$ system. Then 
\begin{eqnarray}
G_0(q')=\frac{1}{\sqrt{\bVec{P}^2+M_{\pi\Sigma}^2}
                 -\sqrt{\bVec{P}^2+W(q')^2}+i\epsilon} \, , 
\label{F14}
\end{eqnarray}
where $\bVec{P}$ is the $\bar K N$ total momentum, and
$W(q')$ is defined by the momentum $\bVec{q}'$  of the $\bar K$ 
in the c.m. frame: 
\begin{eqnarray}
 &&\bVec{P}=\bVec{q}_1+\bVec{q}_2=\bVec{p}_{\pi}+\bVec{p}_{\Sigma}\quad ,
\nonumber\\
 &&W(q')=\sqrt{{\bVec{q}'}^2+ m_{\bar K}^2} +\sqrt{{\bVec{q}'}^2+ m_N^2}\quad .
\label{F15}
\end{eqnarray}
In order to expose the $n \bar K N$ three-body unitarity cut
explicitly, we rewrite Eq.~(\ref{F14}) as
\begin{eqnarray}
G_0(q')=&&\frac{1}{M_{\pi\Sigma}-W(q')+i \epsilon}
\nonumber\\
&&\times\,
\frac{\sqrt{\bVec{P}^2+M_{\pi\Sigma}^2}
                 +\sqrt{\bVec{P}^2+W(q')^2}}
     {M_{\pi\Sigma}+W(q')} \, . 
\label{F16}
\end{eqnarray}
In particular the singular part is given by
\begin{eqnarray}
\frac{1}{M_{\pi\Sigma}-W(q')+i \epsilon}=
\frac{1}{q_0^2-{q'}^2+i \epsilon}\: f(q_0,\: q') \, , 
\label{F17}
\end{eqnarray}
where $q_0$ is defined by
\begin{eqnarray}
 W(q_0) =M_{\pi\Sigma} \, ,
\label{F171}
\end{eqnarray}
and
\begin{eqnarray}
 f(q_0,\: q')^{-1}=[E_1(q_0)&&+E_1(q')]^{-1}
\nonumber\\
 && +[E_2(q_0)+E_2(q')]^{-1} \;\; .
\label{F18}
\end{eqnarray}
Consequently, one finds
\begin{eqnarray}
G_0(q')=&&\frac{1}{q_0^2-{q'}^2+i \epsilon}\: f(q_0,\: q') \qquad
\nonumber\\
&&\times\,
\frac{\sqrt{\bVec{P}^2+M_{\pi\Sigma}^2}
                 +\sqrt{\bVec{P}^2+W(q')^2}}
     {M_{\pi\Sigma}+W(q')} \;\; . 
\label{F185}
\end{eqnarray}
The
c.m. momentum $\bVec{q}'$  of the $\bar K$
is related to the lab momentum
$\bVec{q_2}$ of the nucleon by the relation \cite{FongSucher}
\begin{eqnarray}
 \bVec{q}'=\frac{\epsilon_2\bVec{q}_1-\epsilon_1\bVec{q}_2}
 {\epsilon_1+\epsilon_2} 
 = -\bVec{q}_2+\frac{\epsilon_2}{\epsilon_1+\epsilon_2}  \bVec{P} , 
\label{F19}
\end{eqnarray}
where $\epsilon_i=(E_i+E_i^{cm})/2$,  $(i=1,2)$.
Thereby, in practice,  we change the integral variable  $\bVec{q_2}$ 
in Eq.~(\ref{F2})
to $\bVec{q}'$ and then we can treat the $n \bar K N$ three-body cut 
in Eq.~(\ref{F185}) precisely. 

In the actual calculation the deuteron wave function of the 
Nijmegen soft-core potential Nijm93 \cite{Nijm93} is employed. 
Test calculations performed with the wave function of the CD Bonn
potential \cite{Machleidt} led to practically identical results.
Note that we used both the $S$ and $D$ wave components but the
latter has no visible effect on the considered observables.


\section{Results and Discussion}
\label{sec:results}

\begin{figure}[t]
\includegraphics[width=7cm,angle=-90]{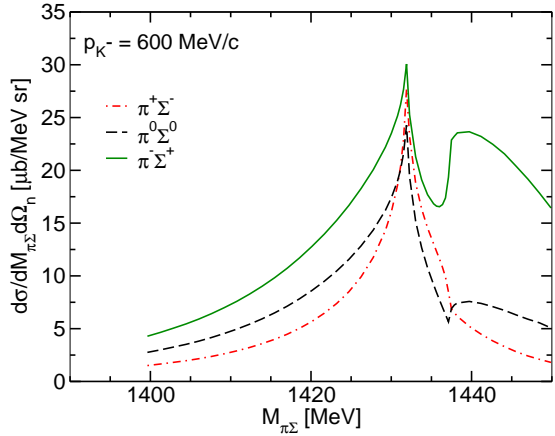}
\caption{$\pi\Sigma$ invariant mass spectrum for the reaction 
$K^- d \to \pi\Sigma n$ at the $K^-$ beam momentum of $600$ MeV/$c$ 
and neutron angle $\theta_n\!=\!0^\circ$.
The $K^- N \to \pi\Sigma$ amplitudes of the J\"ulich model \cite{MG} are used. 
}
\label{f5_1}
\end{figure}
\begin{figure}
\includegraphics[width=7cm,angle=-90]{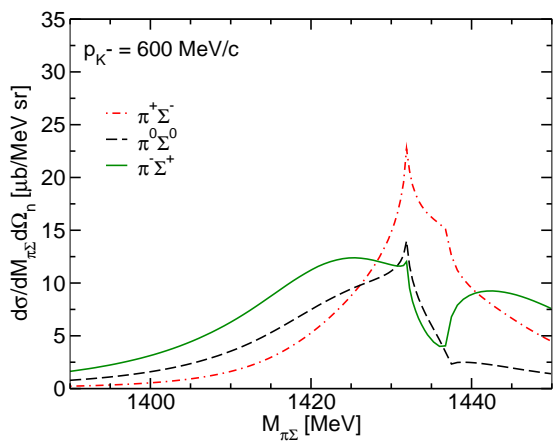}
\caption{$\pi\Sigma$ invariant mass spectrum for the reaction 
$K^- d \to \pi\Sigma n$ at the $K^-$ beam momentum of $600$ MeV/$c$ 
and neutron angle $\theta_n\!=\!0^\circ$.
The $K^- N \to \pi\Sigma$ amplitudes of the Oset-Ramos model \cite{ORB02} are used.
}
\label{f5_2}
\end{figure}

Inclusive cross sections for the reaction $d(K^-, n)\Sigma \pi$
are shown in Figs.~\ref{f5_1} and \ref{f5_2}
where the J\"ulich meson-exchange \cite{MG,Hai11} and the Oset-Ramos chiral 
interaction~\cite{OR98, ORB02} are used for generating the $\bar K N-\pi\Sigma$ 
amplitude, respectively.
We fixed the  $K^-$ beam momentum to $p_{K^-}\!=\!600$ MeV/$c$ and the
neutron angle to $\theta_n\!=\!0^\circ$, considering the kinematics of the J-PARC 
experiment~\cite{jparc} where the neutron is planned to be detected at forward angle.
Taking a glance at the figures, one immediately finds that no clear peaks 
are seen below the $n K^- p$ threshold ($M_{\pi\Sigma} \simeq\!\! 1432$ MeV) for both 
the J\"ulich and the Oset-Ramos potentials. Only for the $\pi^-\Sigma^+ n$ final state 
of the latter interaction (Fig.~\ref{f5_2}) a fairly broad enhancement around 
$M_{\pi\Sigma}=1425$ MeV is visible, however with a shape strongly deformed by the threshold.
Obviously our results are in strong contradiction to the preceding work 
by Jido {\it et al.}~\cite{Jido09} where the same kind of calculation, using the Oset-Ramos 
potential, shows clear peaks below the threshold for all of the three final states. 
In Ref.~\cite{Jido09} lineshapes of the $\pi\Sigma$ invariant mass spectra integrated 
over neutron angles are presented but in Ref.~\cite{jparc} lineshapes limited to 
$\theta_n=0^\circ$ are given by these authors, which exhibit similar peaks to the integrated 
ones. Since these peaks provide the basis of their argument with regard to the $\Lambda (1405)$ 
resonance position, first we want to clarify where this conspicuous difference comes from. 
 
\begin{figure}[t]
\includegraphics[width=7cm,angle=-90]{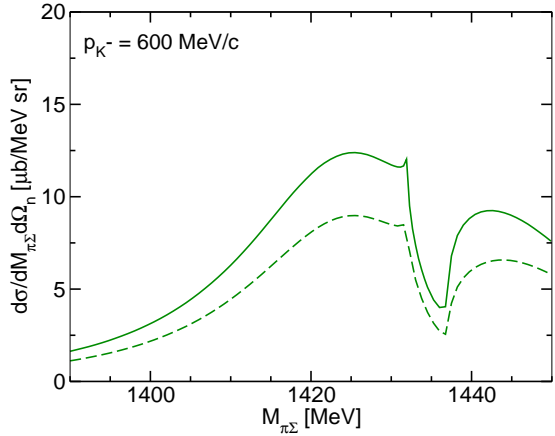}
\caption{$\pi^-\Sigma^+$ invariant mass spectrum for the reaction 
$K^- d \to \pi^-\Sigma^+ n$ at the $K^-$ beam momentum of $600$ MeV/$c$ 
and neutron angle $\theta_n\!=\!0^\circ$. The solid line is the correct 
result while the dashed line is obtained by factorizing the two-body
amplitudes in the loop integral of the two-step process B. 
The $K^- N \to \pi\Sigma$ amplitudes of the Oset-Ramos model \cite{ORB02} are used. 
}
\label{f5_3}
\end{figure}

We start with examining the factorization approximation to the integral
in Eq.(\ref{F2}), which is applied in Ref.~\cite{Jido09}. 
Corresponding results are presented in Fig. \ref{f5_3}. 
This approximation pulls the two amplitudes $t^F$ and $t^I$ out of the 
integral, fixing the momentum variables for these amplitudes to
\begin{eqnarray}
      \bVec{p}_1&=&-\bVec{q}_2\approx  0  \: , \nonumber\\
      \bVec{q}_1&\approx& \; \bVec{p}_{K^-} -\bVec{p}_n   
\label{F21}
\end{eqnarray}
which are the values that give the maximum of the deuteron wavefunction
(see Eq.~(\ref{F3}), and keep in mind that we work in the deuteron rest frame).
Furthermore, the two-body energy $E^I$ of the full off-shell 
 $t$-matrix $t^I$ is approximated by 
\begin{eqnarray}
E^I&=&E_{total}-E_2(q_2)   \nonumber\\
   &=&E_{K^-}+m_d-E_2(q_2) \nonumber\\
   &\approx&E_{K^-}+m_N  
\label{F22}
\end{eqnarray}
(see Eq.~(\ref{F132}) for the definition of $E_2(q_2)$). Then
the second term of the right-hand side of Eq.~(\ref{F2}) is expressed
as 
\begin{eqnarray}
&\sqrt{2} &\int \frac{d^3 q_2}{(2\pi)^3}\,
t^F(\bVec{p}_{\pi}\,\bVec{p}_{\Sigma}\, ,
\bVec{q}_1\bVec{q}_2) \,
G_0 (\bVec{q}_1 \bVec{q}_2)  
\nonumber \\
&&  \hspace{18mm} \times\,
t^I(\bVec{q}_1\,\bVec{p}_n\, ,
\bVec{p}_{K^-}\bVec{p}_1)\, \Phi_d(\bVec{p}) 
\nonumber \\
&& \approx
\sqrt{2} \;\;
t^F_{app} \; t^I_{app}
 \,
\nonumber \\
&&  \hspace{10mm} \times\,
\int \frac{d^3 q_2}{(2\pi)^3}\,
G_0 (\bVec{q}_1 \bVec{q}_2 ) \, \Phi_d(\bVec{p})  \, ,  
\label{F20}
\end{eqnarray}
where $t^F_{app}$ and $t^I_{app}$ are the pertinent amplitudes corresponding 
to the kinematics specified in Eqs.~(\ref{F21}) and (\ref{F22}).
In Fig.~\ref{f5_3} we illustrate the effect of the factorization in the case of the 
final state $\pi^-\Sigma^+ n$. One can see that the magnitude of the cross 
section is reduced by about 30\%, but the lineshape remains practically unchanged.

\begin{figure}[t]
\includegraphics[width=7cm,angle=-90]{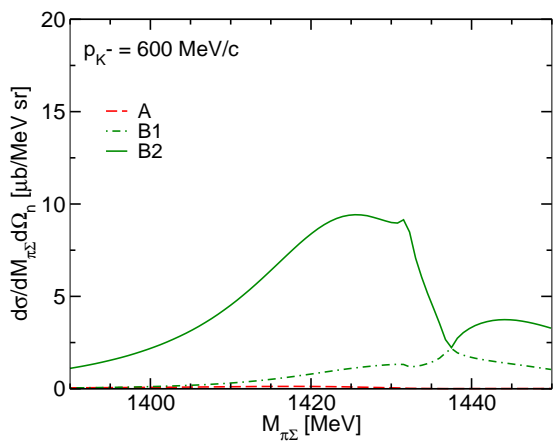}
\caption{$\pi^-\Sigma^+$ invariant mass spectrum for the reaction 
$K^- d \to \pi^-\Sigma^+ n$ at the $K^-$ beam momentum of $600$ MeV/$c$ 
and neutron angle $\theta_n\!=\!0^\circ$. The individual contributions from
the reaction mechanisms A (dashed line), B1 (dash-dotted line), and 
B2 (solid line) are shown based on the factorization approximation. 
The $K^- N \to \pi\Sigma$ amplitudes of the Oset-Ramos model \cite{ORB02} are used.
}
\label{f5_4}
\end{figure}

Before moving to the more crucial approximation adopted in \cite{Jido09}, 
we show individual contributions from the processes A, B1 and B2 
(depicted in Fig.~\ref{f4_1}) where the factorization approximation
is applied for B1 and B2, see Fig.~\ref{f5_4}.
As already mentioned, the cross section for the process A is quite small.
The large momentum of the outgoing neutron, which is directly emitted from 
the deuteron in the case of process A, leads to a tiny value of the deuteron 
wavefunction and suppresses the process
(note that the momentum $\tilde p$ in Eqs.~(\ref{F2}) and (\ref{F3}) is 
$3.9$~fm$^{-1}$ at the $n K^- p$ threshold).  
As is seen in Fig.~\ref{f5_4}, the process B2 yields the main contribution. 
This is due to the fact that the amplitude $t^I (K^- n\rightarrow K^-n)$ 
that enters the process B2 is much larger than $t^I(K^- p\rightarrow \bar K^0 n)$ 
in B1 at $p_{K^-}= 600$ MeV/$c$, something that was already pointed out in 
Ref.~\cite{Jido09}. 
 
Now let us reveal why no clear peaks are seen in our results of the cross 
sections, in contrast to what was shown in Ref.~\cite{Jido09}.
In this reference the authors applied the same approximation as 
introduced to $E^I$ in Eq.~(\ref{F22}) also to the intermediate $\bar K$ energy 
$E_1$ in the propagator $G_0 (\bVec{q}_1 \bVec{q}_2 )$ given in Eq.~(\ref{F131}):
\begin{eqnarray}
E_1&=&E_{total}-E_n-E_2(q_2)  
\nonumber\\
 &\approx &  E_{K^-}+m_N-E_n  
\label{F23}
\end{eqnarray}
(see Eq.~(14) in Ref.~\cite{Jido09}). Then it follows that 
\begin{eqnarray}
G_0(\bVec{q}_1 \bVec{q}_2)
 &=&\frac{1}{E_{total}-E_n-E_2(q_2)-E_1(q_1)+i\epsilon}
\nonumber\\
 &\approx&\frac{1}{E_{K^-}+m_N-E_n-E_1(q_1)+i\epsilon} \:\: .
\label{F24}
\end{eqnarray}
This approximation has a serious impact on the lineshapes of the cross 
section as we will see. Comparing it with the expressions without the 
approximation, Eqs.~(\ref{F13}), (\ref{F14}) and (\ref{F185}),  
one already suspects that it shifts the $n \bar K N$ three-body 
unitarity cut and the $n \bar K N$  threshold position.  
 
\begin{figure}[t]
\includegraphics[width=7cm,angle=-90]{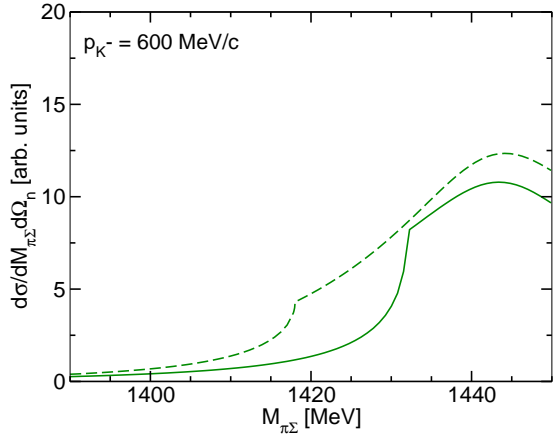}
\caption{$\pi\Sigma$ invariant mass spectrum for the reaction 
$K^- d \to \pi\Sigma n$ at the $K^-$ beam momentum of $600$ MeV/$c$ 
and neutron angle $\theta_n\!=\!0^\circ$. Shown are results based on
the integral over the Green's function alone, cf. Eq.~(\ref{F20}), 
where either the correct expression Eq.~(\ref{F185}) (solid line) or 
the approximation Eq.~(\ref{F24}) (dashed line) are used. 
}
\label{f5_5}
\end{figure}

In order to make the effect of this approximation more transparent we 
work within the factorization approximation (Eq.~(\ref{F20})) and we
consider here the two ingredients that provide the dominant momentum 
dependence in the evaluation of the cross section separately, namely the 
$\bar K N \to \pi \Sigma$ amplitude $t^F_{app}$ and the integral 
$\int d^3 q_2\, G_0 (\bVec{q}_1 \bVec{q}_2 ) \, \Phi_d(\bVec{p}) $.
We focus on the process B2 that yields the overall largest contribution. 
Results based on the assumption that the matrix element
$\langle\bVec{p}_n\vert \langle\bVec{p}_{\pi}\vert
\langle\bVec{p}_{\Sigma}\vert \, T \, \vert \bVec{p}_{K^-}\rangle
\vert\Phi_{d}\rangle\, $
is given solely by the integral over the Green's function 
and the deuteron wave function are presented in Fig.~\ref{f5_5} where 
the solid and dashed lines correspond to the cases without and with the 
approximation described by Eq.~(\ref{F24}), respectively.
One can see that the approximation shifts the $n K^- p$ threshold to
lower energies by an amount of 14~MeV as compared to its actual physical 
value. Furthermore, one realizes that
the integral that enters Eq.~(\ref{F20}) generates a characteristic behavior  
of the cross section at the threshold, in particular a rapid decrease below the threshold,
which comes from the principal-value part of the integral over $G_0 (\bVec{q}_1 \bVec{q}_2 )$.  
 
\begin{figure}[t]
\includegraphics[width=7cm,angle=-90]{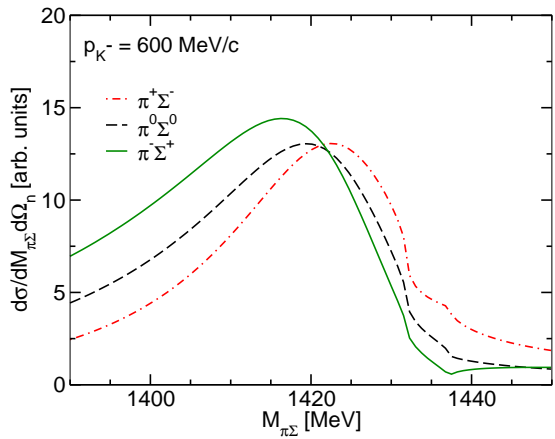}
\caption{$\pi\Sigma$ invariant mass spectrum for the reaction 
$K^- d \to \pi\Sigma n$ at the $K^-$ beam momentum of $600$ MeV/$c$ 
and neutron angle $\theta_n\!=\!0^\circ$. The results are based on using 
only $t^F_{app}$ in Eq.~(\ref{F20}) and considering only the reaction mechanism B2. 
The $K^- N \to \pi\Sigma$ amplitudes of the Oset-Ramos model \cite{ORB02} is used.
}
\label{f5_6}
\end{figure}
\begin{figure}[t]
\includegraphics[width=7cm,angle=-90]{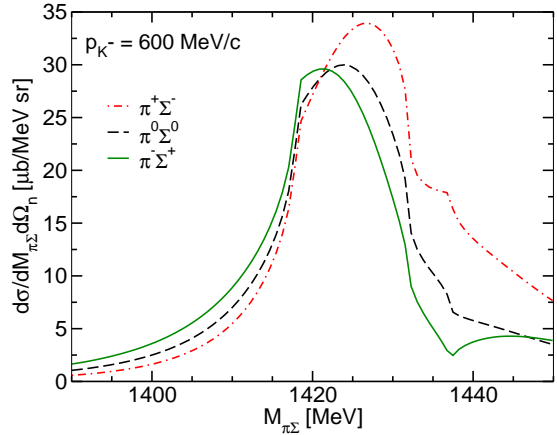}
\caption{$\pi\Sigma$ invariant mass spectrum for the reaction 
$K^- d \to \pi\Sigma n$ at the $K^-$ beam momentum of $600$ MeV/$c$ 
and neutron angle $\theta_n\!=\!0^\circ$. The results are based on the 
approximation Eq.~(\ref{F24}) considering only the reaction mechanism B2. 
The $K^- N \to \pi\Sigma$ amplitudes of the Oset-Ramos model \cite{ORB02} are used.
}
\label{f5_7}
\end{figure}

In Fig.~\ref{f5_6} we display results for the cross sections where the matrix element 
Eq.~(\ref{F20}) is now assumed to be given by the amplitude $t^F_{app}$ alone. 
The cross sections for the three charge states
are displayed, each of which shows a clear peak around $M_{\pi\Sigma}=1420$ MeV.
As expected (and checked by us) 
those lineshapes agree pretty well with the {\it two-body} invariant mass
distributions due to the $K^- p\rightarrow \pi\Sigma$ amplitudes.
Finally, in Fig.~\ref{f5_7}, we plot the cross section based on the full matrix element 
of Eq.~(\ref{F20}) but with the approximation of Eq.~(\ref{F24}) for the Green's
function. (Please note that in the employed factorization approximation this amounts 
practically to the product of the results shown in Figs.~\ref{f5_5} and \ref{f5_6}.)
As already seen and discussed above, the $n K^- p$ threshold is shifted to
lower energies, specifically to $M_{\pi\Sigma}\simeq\!\! 1418$ MeV. 
As a consequence this artificial threshold position is then very close
to the energy where the amplitude $t^F_{app}$ has its peak so that this
approximation generates a huge bump of the cross section just at that
energy. 
Please compare the result for the $\pi^-\Sigma^+ n$ final state in Fig.~\ref{f5_7}
with the solid line in Fig.~\ref{f5_4} where the approximation of Eq.~(\ref{F24}) 
is not made! 

The above considerations strongly suggest that in a precise calculation 
where the three-body ($n \bar K N$) unitarity cut is implemented correctly 
the peaks which are present in the {\it two-body}
amplitude $t^F$, due to the $\Lambda (1405)$, are suppressed by the threshold 
behavior of the Green's function $G_0 (\bVec{q}_1 \bVec{q}_2 )$, so that 
no clear peak (besides a threshold cusp) appears in the corresponding
$\pi \Sigma$ invariant mass spectrum of the three-body final state. 
We believe that this explains the difference between our result and
the one by Jido et al. \cite{Jido09}. In the latter the peak due to the 
$\Lambda (1405)$ in the three-body case is seen at practically the same 
invariant mass as in the two-body amplitude - because approximations are 
applied to the Green's function that shift the opening of the three-body cut
to a lower invariant mass. 

The success of the paper by Jido et al. in stimulating experimental efforts
(and pertinent proposals) results not least from the fact that their calculation 
is roughly in line with data of an old measurement of the $\pi^+ \Sigma^-$ 
invariant mass spectrum for the reaction in question from 1977 \cite{Braun}. 
Those data suggest a peak around $M_{\pi\Sigma}\approx 1425$ MeV - 
and not at the $n K^- p$ threshold! 
In our own calculation within a similar approach, but where now the $n K^- p$ 
unitarity cut is implemented correctly, those data are no longer reproduced. 
However, we would like to emphasize that in a full 
calculation, where all rescattering processes are summed up to infinite 
order as it is the case in Faddeev-type approaches, it is certainly possible
that the structure due to the $\Lambda (1405)$ could survive, after the 
characteristic behavior of $G_0 (\bVec{q}_1 \bVec{q}_2 )$ is smoothened out.
Such a calculation would then not only have the opening of the $n K^- p$ 
channel at the correct location, it would also fulfill exact three-body unitarity,
which is not the case in our study (and also not in Ref.~\cite{Jido09})
where only two-step processes are considered.

\begin{figure}[t]
\includegraphics[width=7cm,angle=-90]{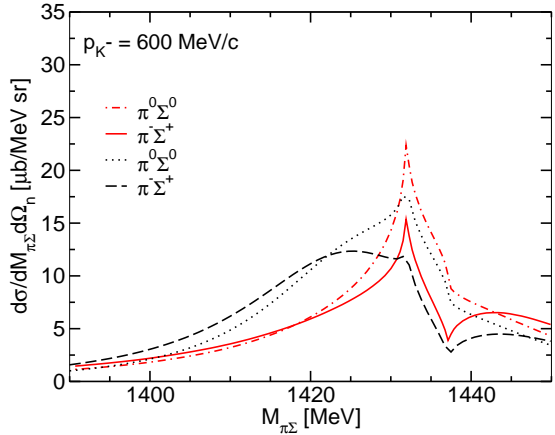}
\caption{$\pi\Sigma$ invariant mass spectrum for the reaction 
$K^- d \to \pi\Sigma n$ at the $K^-$ beam momentum of $600$ MeV/$c$ 
and neutron angle $\theta_n\!=\!0^\circ$.
Comparison of results based on the $K^- N \to \pi\Sigma$ amplitudes 
of the J\"ulich (dash-dotted and solid lines) \cite{MG} 
and the Oset-Ramos (dotted and dashed lines) \cite{ORB02} models.
Only the reaction mechanism B2 is taken into account.
}
\label{f5_8}
\end{figure}

\begin{figure}[t]
\includegraphics[width=7cm,angle=-90]{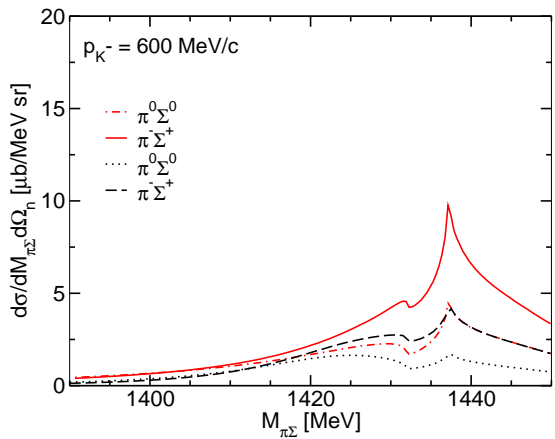}
\caption{$\pi\Sigma$ invariant mass spectrum for the reaction 
$K^- d \to \pi\Sigma n$ at the $K^-$ beam momentum of $600$ MeV/$c$ 
and neutron angle $\theta_n\!=\!0^\circ$.
Comparison of results based on the $K^- N \to \pi\Sigma$ amplitudes 
of the J\"ulich (dash-dotted and solid lines) \cite{MG} 
and the Oset-Ramos (dotted and dashed lines) \cite{ORB02} models.
Only the reaction mechanism B1 is taken into account.
}
\label{f5_9}
\end{figure}

Since within our calculation based on two-step processes, the cross 
sections below the $n K^- p$ threshold turn out to be suppressed by the 
Green's function, we refrain from discussing the results below the threshold, 
i.e. in the $\Lambda$(1405) resonance region, in detail. Rather we focus 
on the differences in the predictions for the $\pi\Sigma$ 
invariant mass spectra at $p_{K^-}\!=\!600$ MeV/$c$
based on the J\"ulich and the Oset-Ramos interactions,
as seen in Figs.~\ref{f5_1} and \ref{f5_2}.
To shed light on this difference let us compare the individual contributions 
from the processes B1 and B2 for the two interactions in question. 
This is done in Figs.~\ref{f5_8} and \ref{f5_9}, exemplary for the final states 
$\pi^0\Sigma^0 n$ and $\pi^-\Sigma^+ n$. 
As is clear from Fig.~\ref{f5_8}, the contributions from the process B2
predicted by those two interactions are very similar above the 
$n K^- p$ threshold. On the other hand, the cross section for 
the process B1 by the Oset-Ramos interaction is much smaller than
that of the J\"ulich potential and amounts to just about 40\% as compared
to the latter at the $n \bar K^0 n$ threshold, cf. Fig.~\ref{f5_9}. 
These two processes (B1 and B2) 
interfere and produce the differences seen between Figs.~\ref{f5_1} and \ref{f5_2}.
We have confirmed that the difference due to B1 above comes from the difference of 
the amplitude $t^I(K^- p\rightarrow \bar K^0 n)$ in the relevant $s$ wave. 
For example, the corresponding $K^- p\rightarrow \bar K^0 n$ (two-body) 
cross section
at $p_{K^-}=600$ MeV/$c$ is 3.62~mb for the J\"ulich interaction while it 
is just 1.57~mb for the Oset-Ramos interaction. 
On the other hand, the $K^- n$ elastic total cross sections at $p_{K^-}=600$ MeV/$c$ 
based on the amplitude $t^I (K^- n\rightarrow K^-n)$ 
that enters into the process B2 are similar for the two interactions:
13.8~mb for the J\"ulich and 13.5~mb for the Oset-Ramos interaction

\begin{table}[h]
\caption{Various $\bar KN$ $s$-wave cross sections in $mb$ for
$p_{K^-} \approx 600$ MeV/c. Results are given for the J\"ulich
\cite{MG} and Oset-Ramos (OR) \cite{OR98} $\bar KN$ interactions 
and two partial wave analyses \cite{ALS,GOP}. 
}
\vskip 0.2cm
\begin{tabular}[t]{ l c c c c }
\hline
channel & \ J\"ulich \ & \ OR \ & Alston \cite{ALS} & Gopal \cite{GOP} \\
\hline
\hline
$K^- p \to K^- p$ & 13.5 & 22.4 & 13.5 & 13.9 \\
$K^- n \to K^- n$ & 13.8 & 13.5 &  6.9 & 7.5 \\
$K^- p \to \bar K^0 n $ & 3.62 & 1.57 &  1.62 & 1.81 \\
\hline
\end{tabular}
\label{Table1}
\end{table}
 
In table \ref{Table1} we summarize the $s$-wave cross sections for
various channels at $p_{K^-} = 600$ MeV/$c$ and compare them with 
results of two partial wave analyses from the 1970s \cite{ALS,GOP}. 
Obviously the predictions of the J\"ulich model agree well with the 
$s$-wave $K^- p$ scattering cross section deduced from empirical 
information but overshoot the other channels, while the
Oset-Ramos interaction is only in line with phenomenology in case 
of the charge-exchange reaction. This may be not too surprising in
view of the fact that both models were primarily designed to 
reproduce the $\bar KN$ data near threshold. 
On the other hand, it is obvious that for a future quantitative 
analysis of the reaction $K^- d \to \pi \Sigma n$, two-body amplitudes
for $K^- n\rightarrow K^-n$ and $K^- p\rightarrow \bar K^0 n$
are required that are fully consistent with the available scattering data.   
Furthermore, one should not forget that at momenta around
$600$ MeV/$c$ higher partial wave could already play a role,
an issue which likewise has to be addressed in a quantitative analysis
of upcoming experimental information. 

\section{Summary}

We investigated the reaction $K^- d \to \pi \Sigma N$ 
taking into account single scattering and the 
two-step process due to $\bar K N \to \pi \Sigma$ rescattering. 
This reaction is considered as a promising candidate for 
exploring the properties of the $\Lambda(1405)$ resonance. 
 
The main aim of our work was to examine the influence of some 
approximations applied in earlier studies of this reaction \cite{Jido09} 
on the results for the $\pi \Sigma$ invariant mass spectrum. 
In particular, in our study we did not use factorization, i.e. the 
$\bar K N \to \bar K N$ and $\bar K N \to \pi \Sigma$ amplitudes 
that enter into the calculation of the two-step process are not pulled
out of the loop integral, and, more importantly, we treated the kinematics 
in the Green's function that appears in the loop integral properly. 
Specifically we made sure that the three-body unitarity cut for the 
intermediate $\bar K NN$ system occurs at the correct (physical) 
threshold. 
In addition we consider different models for the elementary 
$\bar KN$-$\pi \Sigma$ interaction. 

We found that the factorization approximation leads to an overall
reduction in the magnitude of the predicted invariant mass spectrum
in the order of roughly 30\%. However, the lineshape itself remains 
practically unchanged by this approximation. 
On the other hand, the approximation in the kinematics of the Green's 
function, also applied in the works of Jido et al. \cite{Jido09,Jido10}, 
has a rather dramatic impact on the resulting lineshape. 
This approximation shifts the three-body cut due to the opening of
the $nK^-p$ threshold down by roughly 14 MeV from its physical value.
It then coincides practically with the peak value of the
elementary $\bar KN$-$\pi \Sigma$ amplitude that corresponds to the
$\Lambda$(1405) resonance and, consequently, the resulting lineshape
exhibits a strong enhancement at a $\pi \Sigma$ invariant mass of
around 1426 MeV. In contrast, a calculation where the
$nK^-p$ cut is taken into account precisely leads to a $\pi \Sigma$ 
invariant mass spectrum that has a pronounced peak around 1435 MeV,
i.e. at the opening of the $nK^-p$ channel. Indeed the peak is 
nothing else than a threshold effect (cusp). 
In that calculation the contribution of the rescattering process
$\bar KN$-$\pi \Sigma$ around 1426 MeV, where this amplitude has
its maximum, is already significantly suppressed by the deuteron
wave function and the fall-off of the $nK^-p$ Green's function.
Thus, only a rather broad structure is visible in the spectrum
for the case of the $\bar KN$-$\pi \Sigma$ generated from the
Oset-Ramos interaction whereas for the J\"ulich $\bar KN$ model
there is no direct sign at all of the $\Lambda$(1405) resonance. 

Interestingly, existing data on the $\pi^+ \Sigma^-$ invariant mass spectrum 
from the reaction $K^- d \to \pi \Sigma N$ \cite{Braun} seem to suggest a 
peak around $M_{\pi\Sigma}\approx 1425$ MeV - and not at the $n K^- p$ 
threshold! Should that be confirmed in the planned measurements at J-PARC
it would be certainly a sign for the inadequacy of the approach adopted so 
far in the pertinent investigations \cite{Jido09,Jido10}. And it would
be a strong hint that one should rather rely on Faddeev-type approaches 
where all rescattering processes can be summed up to 
infinite order. Then the structure in the two-body amplitudes 
corresponding to the $\Lambda$(1405) resonance can be generated within
the three-body context in a consistent way. Moreover, exact three-body 
unitarity can be automatically fulfilled. Such a calculation is beyond
the scope of the present investigation but we intend to 
address this issue in a future study. 

\acknowledgments{
We would like to thank A. Ramos for providing us with the code
for the potential of Refs.~\cite{OR98,ORB02}. 
K.M. thanks Walter Gl\"ockle for valuable discussions on
the basic equations for the $K^- d$ reaction.
}

\end{document}